\documentclass[conference]{IEEEtran}
\usepackage{graphicx}%------------------------------added by ZY
\usepackage{subfigure}
\usepackage{float}%---------------------------------added by ZY
\usepackage{algorithm}%-----------------------------added by ZY
\usepackage{algorithmic}%---------------------------added by ZY
\usepackage{flushend}%------------------------------added by ZY
\usepackage{mathrsfs}%------------------------------added by ZY
\usepackage{amsfonts}
\usepackage{array}
\usepackage{amsmath,amssymb,amsfonts,mathrsfs}
\usepackage{longtable}
\usepackage{rotating}
\usepackage{multirow}
\usepackage{epstopdf}
\usepackage{flushend}%%%%%%%%%%%%%%%%%%%%%%%%%%%%%%%%%
\usepackage{cite}
\usepackage{float}
\usepackage[marginal]{footmisc}
\usepackage{stfloats}
\usepackage[colorlinks,
            linkcolor=black,
            anchorcolor=black,
            citecolor=black]{hyperref}

\renewcommand{\baselinestretch}{1}

\def\BibTeX{{\rm B\kern-.05em{\sc i\kern-.025em b}\kern-.08em
    T\kern-.1667em\lower.7ex\hbox{E}\kern-.125emX}}

\renewcommand{\baselinestretch}{1}
\newtheorem{definition}{Definition}

\begin{document}

%\title{Dynamic Social Awareness D2D-Aided Cooperative MEC for Low Latency IoT}
%\title{Price-Based Resource Management for Intelligent Reflecting Surface Aided Communications: A Stackelberg Game Approach}
\title{A Stackelberg Game Approach to Resource Allocation for Intelligent Reflecting Surface Aided Communications}
\author{\IEEEauthorblockN{Yulan Gao\textsuperscript{1,2}, Chao Yong\textsuperscript{1}, Zehui Xiong\textsuperscript{2}, Dusit Niyato\textsuperscript{2},  Yue Xiao\textsuperscript{1}, Jun Zhao\textsuperscript{2}}

\IEEEauthorblockA{\textsuperscript{1}{The National Key Laboratory of Science
and Technology on Communications} \\
{University of Electronic
Science and Technology of China, Chengdu 611731, China }\\
{email: xiaoyue@uestc.edu.cn}}

\IEEEauthorblockA{\textsuperscript{2}{The School of Computer Science and Engineering,}
{Nanyang Technological University, Singapore  639798}}
}

\markboth{~}
\maketitle
\vspace{-3.3em}
\maketitle
\begin{abstract}
%Intelligent reflecting surface (IRS) exploits large reflection elements to proactively steer the incident radio-frequency wave towards destination terminals (DTs), which is a promising solution to build a programmable wireless environment for 6G systems.
It is known that the capacity of the intelligent reflecting surface (IRS) aided cellular network can be effectively improved by reflecting the incident signals from the transmitter in a low-cost passive reflecting way.
In this paper, we study the adoption of an IRS for downlink multi-user communication from a multi-antenna base station (BS).
Nevertheless, in the actual network operation, the IRS operator can be selfish or have its own objectives due to competing/limited resources as well as deployment/maintenance cost.
Therefore, in this paper, we develop a Stackelbeg game model to analyze the interaction between the BS and the IRS operator.
Specifically, different from the existing studies on IRS that merely focus on tuning the reflection coefficient of all the reflection elements, we consider the  reflection resource (elements) management, which can be realized via trigger module selection under our proposed IRS architecture that all the reflection elements are partially controlled by independent switches of controller.
A Stackelberg game-based alternating direction method of multipliers (ADMM) is proposed to jointly optimize the transmit beamforming at the BS and the passive beamforming of the triggered reflection modules.
Numerical examples are presented to verify the proposed studies.
It is shown that the proposed scheme is effective in the utilities of both the BS and IRS.
\end{abstract}

\IEEEpeerreviewmaketitle

\section{Introduction}
\IEEEPARstart{B}{y enabling} the intelligent reflecting surface (IRS) to the wireless systems, the IRS-aided wireless system recently has attracted significant interest due to its potential to further improve the system capacity and spectral efficiency \cite{Wu2019Beamforming, Huang2019Reconfigurable, Zenzo2019Wireless, Zenzo2019Smart}.
Specifically, IRS exploits large reflection elements to proactively steer the incident radio-frequency wave towards destination terminals \cite{Basar2019Wireless}, which is a promising solution to build a programmable wireless environment for 6G systems \cite{Hu2019Reconfigurable, Wu2019Towards}.
Thereby, the fine-grained three-dimensional reflecting beamforming can be achieved without the need of any transmit radio frequency (RF) chain \cite{Han2019Large}.
%This approach can achieve fine-grained three-dimensional reflecting beamforming without the need of any transmit radio-frequency (RF) chain

\subsection{Related Work}
The IRS-aided wireless systems refer to the scenario that a large number of software-controlled reflection elements  with adjustable phase shifts for reflecting the incident signal.
As such,  the phase shifts of all reflection elements can be tuned adaptively according to the state of networks, e.g., the channel conditions and the incident angle of the signal by the base station (BS).
%Notably, different from the conventional half and full-duplex modes, in IRS-aided communication,
It is commonly believed that the propagation environment can be improved without incurring additional noise at the reflector elements.
Currently, major communication field researchers are actively involved in the research of IRS-aided communications \cite{Gao2020Reconfigurable, Wu2019Intelligentjournal,Guo2019Weighted, Pan2019Intelligent, Bjornson2019Intelligent, Zenzo2019Reconfigurable}.
For example,  \cite{Zenzo2019Reconfigurable} summarized the main communication applications and competitive advantages of the IRS technology.
In the spirit of these works, a vast corpus of literature focused on optimizing active-passive beamforming for unilateral spectral efficiency maximization subject to power constraint.
For instance, \cite{Guo2019Weighted} proposed a fractional programming based alternating optimization approach to maximize the weighted SE in IRS-aided MISO downlink communication systems.
In particular, three assumptions for the feasible set of reflection coefficient  were consider at IRS, including the ideal reflection coefficient constrained by peak-power, continuous phase shifter, and discrete phase shifter.
Meantime, in MISO wireless systems, the problem of minimizing the total transmit power at the access point was considered to
energy-efficient active-passive beamforming \cite{Wu2019Beamforming,Wu2019Intelligentjournal}.
\cite{Wu2019Beamforming} formulated and solved the total transmit power minimization problem by joint active-passive beamforming design, subject to the signal-to-interference-plus-noise ratio (SINR) constraints, where each reflection element is a continuous phase shifter.
Along this direction, considering the discrete reflect phase shifts at the IRS,  the same optimization problem was further studied in \cite{Wu2019Intelligentjournal}.
Notably, the aforementioned studies for IRS-aided communications were based on the premise of ignoring the power consumption at IRS.
In contrast, in \cite{Huang2019Reconfigurable},  an energy efficiency (EE) maximization problem was investigated by developing a realistic IRS power consumption model, where IRS power consumption relies on the type and the resolution of meta-element.

\subsection{ Motivation and Contributions}

The above resource allocation works address the joint transmit beamforming and phase shift optimization problem in IRS-aided communication systems.
These works assume that IRS operators are all selfless, and will always participate in the cooperative transmission despite their own energy consumption/maintanence cost \cite{Huang2019Reconfigurable} and profits.
However, this assumption becomes unrealistic in practice, due to the advances in intelligent communication and the shrinking resources.
In other words, if an IRS operator cannot benefit from the participation, it will not join in the cooperative communication.
Moreover, the  common assumption in the existing studies for IRS-aided communications is that all the reflection elements are used to reflect the incident signal, i.e., adjusting reflecting coefficient of each meta-element simultaneously each time.
However, along with the use of a large number of high-resolution reflection elements, especially with continuous phase shifters, triggering all the reflection elements every time may result in significant  power consumption.
Moreover, the hardware support for the  IRS implementation is the use of a large number of tunable metasurfaces.
Specifically, the tunability feature can be realized by introducing mixed-signal integrated circuits (ICs) or diodes/varactors, which can  vary both the resistance and reactance, offering complete
local control over the complex surface impedance \cite{Liu2019Intelligent,Tan2018Enabling, Li2017Electromagnetic, Zenzo2019Smart}.
According to the IRS power consumption model presented in \cite{Huang2019Reconfigurable} and the hardware support, triggering the entire IRS not only incurs increased power consumption, but also entails the increased latency of adjusting phase-shift and accelerates equipment depreciation.
Therefore, realizing  reflection resource management is significantly important for IRS-aided communications.
In this paper, for IRS-aided multiuser multiple-input single-output (MISO) systems, we consider the resource allocation problem in which an IRS operator serves the BS and prices the triggered reflection module.
%We focus on the pricing mechanism that can be applied by the IRS operator to manage and control
The problem is formulated as a Stackelberg game, in which the IRS operator decides the price for the trigger reflection modules.

The contributions of this paper are summarized as follows:
\begin{itemize}
\item{For the first time,  a modular architecture of IRS is proposed that  divides all the reflection elements into multiple modules which can be independently controlled by parallel switches.  In order to avoid signal loss due to excessive scattering, we assume that each module contains multiple reflection elements, i.e., the size of each module is larger than the incident signal wavelength, since the unit meta-element size is subwavelength \cite{Liu2019Intelligent}.
    As mentioned in \cite{Wu2019Towards},  the IRS is programmatically controlled by the controller, and hence,  from an operational standpoint, independent module triggering can be implemented easily.
    Therefore, the proposed architecture of IRS allows the realization of the  reflection resource management, since each module is independently controlled by its switch.}

\item{Based on the proposed modular architecture of IRS, this paper proposes a new price-based resource allocation scheme for both the BS and IRS.
    Furthermore, the Stackelberg game is formulated to maximize the individual revenue of the BS and IRS for the proposed price-based resource allocation.
    Since the entire game is a non-convex mixed-integer problem, which is even hard to solve in a centralized way, the problem is transformed into a convex problem by introducing the mixed row block $\ell_{1,2}\text{-norm}$ \cite{Mehanna2013Joint}, which yields a suitable semidefinite relaxation.
    To solve this problem, we apply a Stackelberg game-based alternating direction method of multipliers (ADMM) to identify the price, trigger module subsets, and subsequently both the transmit power allocation and the corresponding passive beamforming.
    }
\end{itemize}

The rest of this paper is organized as follows. In Section II, we introduce the system model and formulate the Stackelberg game problem.
Section III investigates the Stackelberg game-based ADMM algorithm,  the optimal price and active-passive beamforming.
Simulation results are provided in Section IV. In Section V, we draw our main conclusions.

\section{System Model and Problem Formulation}
\subsection{Signal Model}
Consider the downlink communication between a BS equipped with $M$ antennas and $K$ single-antenna mobile users. The communication takes place via an IRS with $S$ reflection modules, and each module consisting $N$ reflection elements, and thus,  the total reflection elements of IRS is $SN.$
Define ${\cal K}:=\{1, 2, \ldots, K\},$ ${\cal S}:=\{1, 2, \ldots, S\},$ and ${\cal I}=\{1, 2, \ldots, (SN)\}$ as the index sets of users, the reflection modules, and the reflection elements, respectively.
%The direct signal path between the BS and the mobile users is neglected due to unfavorable propagation conditions.
Let ${\mathbf H}_{0,s}\in{\mathbb C}^{N\times M}$ be the channel matrix from the BS to the $s\text{th}$ reflection module of IRS, ${\mathbf g}_{s,k}\in{\mathbb C}^{N\times 1}$ be the channel vector from the $s\text{th}$ reflection module of the IRS to user $k.$
The direct channel for the BS to user $k$ is denoted as $h_{d,k}\in{\mathbb C}^{M\times 1}.$
Denote by $\phi_i, \forall i\in{\cal I}$ the $i\text{th}$ reflection element of the IRS.
Let ${\pmb\Phi}=\text{diag}\{{\pmb\Phi}_1, {\pmb\Phi}_2, \ldots, {\pmb\Phi}_S\}\in{\mathbb C}^{(SN)\times(SN)},$ where ${\pmb\Phi}_s=\text{diag}[\phi_{(s-1)N+1}, \phi_{(s-1)N+2}, \ldots, \phi_{sN}]\in{\mathbb C}^{N\times N}.$
Define ${\pmb\phi}=[({\pmb\phi}_1)^T,({\pmb\phi}_2)^T,\ldots, ({\pmb\phi}_S)^T ]^T\in{\mathbb C}^{(SN)\times 1}, $  where ${\pmb\phi}_s=[({\phi}_{(s-1)N+1})^{\dag}, ({\phi}_{(s-1)N+2})^{\dag}, \ldots, ({\phi}_{sN})^{\dag}]^T\in{\mathbb C}^{N\times 1}.$

We assume that all the reflection modules of IRS can potentially join the cooperative communication, then, the channel matrix from the BS to the IRS and the IRS to user $k$ respectively are
\begin{equation}\label{eq:1}
\begin{aligned}
{\mathbf H}&=\left[\left({\mathbf H}_{0,1}\right)^T, \left({\mathbf H}_{0,2}\right)^T, \ldots, \left(  {\mathbf H}_{0,S}\right)^T   \right]^T\in{\mathbb C}^{(SN)\times M}\\
{\mathbf g}_k&=\left[ ({\mathbf g}_{1,k})^T, ({\mathbf g}_{2,k})^T, \ldots, ({\mathbf g}_{S,k})^T \right]^T\in{\mathbb C}^{(SN)\times 1}, \forall k\in{\cal K}.
\end{aligned}
\end{equation}
The SINR for user $k$, which is denoted by $\gamma_k$ can be computed by
\begin{equation}\label{eq:2}
\gamma_k=\frac{\left(\left|{\mathbf h}_{d,k}^{\dag}+{\mathbf g}_{k}^{\dag}{\pmb\Phi}{\mathbf H}\right){\mathbf w}_k \right|^2}
{\sum_{j\neq k}^K \left|({\mathbf h}_{d,k}^{\dag}+{\mathbf g}_k^{\dag}{\pmb\Phi}{\mathbf H}){\mathbf w}_j  \right|^2+\sigma^2},
\end{equation}
where ${\mathbf w}_k\in{\mathbb C}^{M\times 1}$ is the transmit beamforming vector for user $k.$

The utility function of the BS is given by
\begin{equation}\label{eq:3}
U=\sum_{k=1}^K \log_2\left( 1+\gamma_k\right)-{ r}||{\pmb\Phi}||_{0,2},
\end{equation}
where ${ r}>0$ is the price to the IRS for providing $||{\pmb\Phi}||_{0,2}$ reflection modules.
Moreover, $||{\pmb\Phi}||_{0,2}\triangleq \left|\left\{s: ||{\pmb\Phi}_s||_2\neq 0  \right\}\right|,$
where ${\pmb\Phi}_s\in{\mathbb C}^{N\times N}$ denotes the $s\text{th}$ diagonal block of matrix ${\pmb\Phi}, s=1, 2, \ldots, S.$ The $\ell_{0,2}-$norm is the number of nonzero diagonal blocks of $\pmb\Phi.$
It is possible to replace any sparsity inducing norm regularization without changing the regularization properties of the problem \cite{Mehanna2013Joint}. We will use the convex $\ell_{1,2}-$norm as a group-sparsity inducing regularization to replace the non-convex $\ell_{0,1}-$norm in (\ref{eq:3}), and the $\ell_{1, 2}-$norm is defined as
\begin{equation}\label{eq:5}
||{\pmb\Phi}||_{1,2}\triangleq \sum_{s=1}^S||{\pmb\Phi}_s||_2.
\end{equation}
Consequently, the utility function of the BS is expressed as
\begin{equation}\label{eq:6}
U=\sum_{k=1}^K \log_2(1+\text{SINR}_k)-{ r}\alpha\sum_{s=1}^S||{\pmb\Phi}_s||_2,
\end{equation}
where balance parameter $\alpha>0$.
Accordingly, the utility of the IRS is defined as the revenues received from the BS, shown as
\begin{equation}\label{eq:7}
V={ r}\alpha\sum_{s=1}^S||{\pmb\Phi}_s||_2.
\end{equation}
\subsection{Stackelberg Game Formulation}
Based on the above discussion, the problem can be formulated as a Stackelbeg game, where the IRS is the leader and the BS is the follower. In a Stackelberg game, the leader selects its strategy to optimize its utility first and then the follower move to optimize its utility based on the leader's startegy.
In particular, here the IRS adjusts the price, ${\mathbf r}$, as the strategy, to maximize its utility.
Thus, the objective of the IRS is to solve the following problem ({\bf L-Problem}):
\begin{equation}\label{eq:8}
\begin{aligned}
\max_{{ r}}~ &V={ r}\alpha\sum_{s=1}^S||{\pmb\Phi}_s||_2\\
\text{s.t.~}& { r}>0.
\end{aligned}
\end{equation}
In response to the action of the IRS (leader), the BS (follower) chooses the best trigger reflection modules, and decides the passitve beamforming of the selected reflection modules and the transmit beamforming at the BS.
The problem of obtaining the optimal strategy for the BS (follower) can be formulated as follows:
\begin{equation}\label{eq:9}
\begin{aligned}
\text{\bf F-Problem}~\max_{{\mathbf w}_k, {\pmb\Phi}}~ &U=\sum_{k=1}^K\log_2(1+\text{SINR}_k)-{\mathbf r}\alpha\sum_{s=1}^S||{\pmb\Phi}_s||_2\\
\text{s.t.~}& \sum_{k=1}^K ||{\mathbf w}_k||_2^2\leq p^{\max}\\
&|\phi_i|\leq 1, \forall i=1,2, \ldots, (SN).
\end{aligned}
\end{equation}

For the proposed Stackelberg game, the Stackelberg game equilibrium (SE) is defined as follows.
\begin{definition}
Define $ {\mathbf W}=[{\mathbf w}_1, {\mathbf w}_2, \ldots,{\mathbf w}_K]\in{\mathbb C}^{M\times K}.$
Let ${ r}^{*}$ be a solution of problem (\ref{eq:8}) and $({\mathbf W}^{*}, {\pmb\Phi}^{*})$ be a solution for problem (\ref{eq:9}). Then, the point $({ r}^{*}, {\mathbf W}^{*}, {\pmb\Phi}^{*})$ is the Stackelberg equilibrium for the proposed Stackelberg game if for any $({ r}, {\mathbf W}, {\pmb\Phi})$, the following conditions are satisfied:
\begin{equation}\label{eq:10}
\begin{aligned}
U({ r}^{*}, {\mathbf W}^{*}, {\pmb\Phi}^{*})&\geq U({ r}^{*}, {\mathbf W}, {\pmb\Phi})\\
V({ r}^{*}, {\mathbf W}^{*}, {\pmb\Phi}^{*})&\geq V({ r}, {\mathbf W}^{*}, {\pmb\Phi}^{*}).
\end{aligned}
\end{equation}
\end{definition}

\section{Game Analysis}
In the proposed game, both at the BS's and the IRS's side, since there is only one player, the best response of the BS and IRS can be readily obtained by solving {\bf F-Problem} and {\bf L-Problem}, respectively.
For the proposed game, the SE can be obtained as follows: For a given ${ r},$ {\bf F-Problem } (\ref{eq:9}) is solved first. Then, with the obtained best response functions $({ W}^{*}, {\pmb\Phi}^{*})$ of the BS, we solve {\bf L-Problem} (\ref{eq:8}) for the optimal price ${ r}^{*}.$

\subsection{Strategy Analysis for the BS}
If we denote the price for serving the BS as ${ r}.$
The follower problem is
\begin{equation}\label{eq:11}
\begin{aligned}
\text{(U0)}~\max_{{\mathbf w}_k, {\pmb\Phi}}~ &U=\sum_{k=1}^K\log_2(1+\gamma_k)-{ r}\alpha\sum_{s=1}^S||{\pmb\Phi}_s||_2\\
\text{s.t.~}& \sum_{k=1}^K ||{\mathbf w}_k||_2^2\leq p^{\max}\\
&|\phi_i|\leq 1, \forall i=1,2, \ldots, (SN).
\end{aligned}
\end{equation}

To tackle the logarithm in the objective function of (\ref{eq:11}), we apply the Lagrangian dual transform.
Then, (U0) can be equivalently written as
\begin{equation}\label{eq:12}
\begin{aligned}
\max_{{\mathbf W}, {\pmb\Phi}}~\sum_{k=1}^K\log_2\left(1+\alpha_k\right)&-\sum_{k=1}^K\alpha_k
+\sum_{k=1}^K\frac{(1+\alpha_k)\gamma_k}{1+\gamma_k}\\
&-{ r}\alpha\sum_{s=1}^S||{\pmb\Phi}_s||_2.
\end{aligned}
\end{equation}
In (\ref{eq:12}), when ${\mathbf W}$ and ${\pmb\Phi}$ hold fixed, the optimal $\alpha_k$ is
\begin{equation}\label{eq:13}
\alpha_k^{*}=\gamma_k, \forall k\in{\cal K}.
\end{equation}
Then, for a given price ${ r}$ and  a fixed $\{\alpha_k\}_{k\in{\cal K}},$ optimizing ${\mathbf W}$ and ${\pmb\Phi}$ is reduced to
\begin{equation}\label{eq:14}
\begin{aligned}
\text{(U0-1)~}\max_{{\mathbf W}, {\pmb\Phi}} \sum_{k=1}^K\frac{\tilde{\alpha}_k\gamma_k}{1+\gamma_k}-{ r}\alpha\sum_{s=1}^S||{\pmb\Phi}_s||_2,
\end{aligned}
\end{equation}
where $\tilde{\alpha}_k=1+\alpha_k.$

\subsubsection{Transmit Beamforming}
In the following, we investigate how to find a better beamforming matrix ${\mathbf W}$ given fixed ${\pmb\Phi}$ for (\ref{eq:14}). Denote the combined channel for user $k$ by
\begin{equation}\label{eq:15}
{\mathbf h}_k^{\dag}={\mathbf h}_{d,k}^{\dag}+{\mathbf g}_k^{\dag}{\pmb\Phi}{\mathbf H}, \forall k\in{\cal K.}
\end{equation}
Then, the SINR $\gamma_k$ in (\ref{eq:2}) is given by
\begin{equation}\label{eq:16}
\gamma_k=\frac{|{\mathbf h}_k^{\dag}{\mathbf w}_k|^2}
{\sum_{j\neq k}^K|{\mathbf h}_k{\mathbf w}_j|^2+\sigma^2}.
\end{equation}
Using $\gamma_k$ in (\ref{eq:16}), the objective function of (\ref{eq:14}) is written as a function of ${\mathbf W}:$
\begin{equation}\label{eq:17}
\begin{aligned}
\sum_{k=1}^K\frac{\tilde{\alpha}_k\gamma_k}{1+\gamma_k}-&{ r}\alpha\sum_{s=1}^S||{\pmb\Phi}_s||_2
=\\&\sum_{k=1}^K\frac{\tilde{\alpha}_k|{\mathbf h}_k^{\dag}{\mathbf w}_k|^2}
{\sum_{j=1}^K|{\mathbf h}_k^{\dag}{\mathbf w}_j|^2+\sigma^2}-{ r}\alpha\sum_{s=1}^S||{\pmb\Phi}_s||_2.
\end{aligned}
\end{equation}
Thus, for given ${ r},$ $\{{\alpha}_k\}_{k\in{\cal K}},$ and ${\pmb\Phi}$, optimizing ${\mathbf W}$ becomes
\begin{equation}\label{eq:18}
\begin{aligned}
\text{(U1-1)}\max_{\mathbf W}& \sum_{k=1}^K\frac{\tilde{\alpha}_k|{\mathbf h}_k^{\dag}{\mathbf w}_k|^2}
{\sum_{j=1}^K|{\mathbf h}_k^{\dag}{\mathbf w}_j|^2+\sigma^2}\\
\text{s.t.~}&\sum_{k=1}^K||{\mathbf w}_k||_2^2\leq p^{\max}.
\end{aligned}
\end{equation}
Using quadratic transform, the objective function of (U1-1) is reformulated as
\begin{equation}\label{eq:19}
\begin{aligned}
\sum_{k=1}^K\frac{\tilde{\alpha}_k|{\mathbf h}_k^{\dag}{\mathbf w}_k|^2}
{\sum_{j=1}^K|{\mathbf h}_k^{\dag}{\mathbf w}_j|^2+\sigma^2}
&=\sum_{k=1}^K2\sqrt{\tilde{\alpha}_k}\text{Re}\left\{ \beta_k^{\ddag}{\mathbf h}_k^{\dag}{\mathbf w}_k \right\}\\
&-\sum_{k=1}^K|\beta_k|^2\left(\sum_{j=1}^K|{\mathbf h}_k^{\dag}{\mathbf w}_j|^2+\sigma^2  \right)
\end{aligned}
\end{equation}
where $(\cdot)^{\ddag}$ denotes the conjugate. $\beta_k\in{\mathbb C}$ is the auxiliary variable. Then, solving problem (U1-1) over ${\mathbf W}$ is equivalent to solving the following problem over ${\mathbf W}$ and ${\pmb\beta}=[\beta_1, \ldots, \beta_K]^T\in{\mathbb C}^{K\times 1}:$
\begin{equation}\label{eq:20}
\begin{aligned}
\text{(U1-2)}~\max_{{\mathbf W}, {\pmb\beta}}~&~\sum_{k=1}^K2\sqrt{\tilde{\alpha}_k}\text{Re}\left\{ \beta_k^{\ddag}{\mathbf h}_k^{\dag}{\mathbf w}_k \right\}\\
&-\sum_{k=1}^K|\beta_k|^2\left(\sum_{j=1}^K|{\mathbf h}_k^{\dag}{\mathbf w}_j|^2+\sigma^2  \right)\\
\text{s.t.~}&\sum_{k=1}^K||{\mathbf w}_k||_2^2\leq p^{\max}.
\end{aligned}
\end{equation}
The optimal $\beta_k$ for a given ${\mathbf W}$ is
\begin{equation}\label{eq:21}
\beta_k^{*}=\frac{\sqrt{\tilde{\alpha}_k}{\mathbf h}_k^{\dag}{\mathbf w}_k}
{\sum_{j=1}^K|{\mathbf h}_k^{\dag}{\mathbf w}_j|^2+\sigma^2}.
\end{equation}
Then, fixing ${\pmb\beta},$ the optimal ${\mathbf w}_k$ is
\begin{equation}\label{eq:22}
{\mathbf w}_k^{*}=\sqrt{\tilde{\alpha}_k}\beta_k\left(\lambda_0{\mathbf I}_M+\sum_{j=1}^K|\beta_j|^2{\mathbf h}_j{\mathbf h}_j^{\dag}\right)^{-1}{\mathbf h}_k,
\end{equation}
where $\lambda_0$ is the dual variable introduced for the power constraint, which is optimally determined by
\begin{equation}\label{eq:23}
\lambda_0^{*}=\max\left\{0, p^{\max}-\sum_{k=1}^K||{\mathbf w}_k||_2^2 \right\}.
\end{equation}
\subsubsection{Optimizing Reflection Response Matrix ${\pmb\Phi}$}
Optimize ${\pmb\Phi}$ in (U0-1) given fixed pricing ${\mathbf r},$ $\{\alpha_k\}_{k\in{\cal K}},$ and ${\mathbf W}.$
Using $\gamma_k$ defined in (\ref{eq:2}), the objective function of (U0-1) is expressed as a function of ${\pmb\Phi}$:
\begin{equation}\label{eq:24}
\sum_{k=1}^K\frac{\tilde{\alpha}_k|({\mathbf h}_{d,k}^{\dag}+{\mathbf g}_k^{\dag}{\pmb\Phi}(\mathbf H)){\mathbf w}_k|^2}
{\sum_{j=1}^K|({\mathbf h}_{d,k}^{\dag}+{\mathbf g}_k^{\dag}{\pmb\Phi}{\mathbf H}){\mathbf w}_j|^2+\sigma^2}-{ r}\alpha\sum_{s=1}^S||{\pmb\Phi}_s||_2
\end{equation}
Define ${\mathbf a}_{j,k}=\text{diag}\{{\mathbf g}_k^{\dag}\}{\mathbf H}{\mathbf w}_j, b_{j,k}={\mathbf h}_{d,k}^{\dag}{\mathbf w}_j,
\forall k,j=1, 2, \ldots, K.$
Combining with the definition of ${\pmb\phi}$, (\ref{eq:24}) can be rewritten as
\begin{equation}\label{eq:25}
\sum_{k=1}^K\frac{\tilde{\alpha}_k|b_{k,k}+{\pmb\phi}^{\dag}{\mathbf a}_{k,k}|^2}
{\sum_{j=1}^K|b_{j,k}+{\pmb\phi}^{\dag}{\mathbf a}_{j,k}|^2+\sigma^2}-{ r}\alpha\sum_{s=1}^S||{\pmb\Phi}_s||_2
\end{equation}
Note that $\sum_{s=1}^S||{\pmb\Phi}_s||_2=\sum_{s=1}^S||{\pmb\phi}_s||_2,$ optimizing ${\pmb\phi}$ can be represented as follows:
\begin{equation}\label{eq:26}
\begin{aligned}
\text{(U2-1)~}\max_{{\pmb\phi}}~&\sum_{k=1}^K\frac{\tilde{\alpha}_k|b_{k,k}+{\pmb\phi}^{\dag}{\mathbf a}_{k,k}|^2} {\sum_{j=1}^K|b_{j,k}+{\pmb\phi}^{\dag}{\mathbf a}_{j,k}|^2+\sigma^2}-{ r}\alpha\sum_{s=1}^S||{\pmb\phi}_s||_2\\
\text{s.t.~} &{\pmb\phi}^{\dag}{\mathbf e}_i{\mathbf e}_i^{\dag}{\pmb\phi}\leq 1, \forall i=1, 2, \ldots, (SN).
\end{aligned}
\end{equation}
Based on the quadratic transform, the new objective function of (U2-1) is
\begin{equation}\label{eq:27}
\begin{aligned}
\sum_{k=1}^K&2\sqrt{\tilde{\alpha}_k}\text{Re}\left\{ \epsilon_k^{\ddag}{\pmb\phi}^{\dag}{\mathbf a}_{k,k} +\epsilon_k^{\ddag}b_{k,k}\right\}-\sum_{k=1}^K|\epsilon_k|^2\\
&\times\left( \sum_{j=1}^K|b_{j,k}+{\pmb\phi}^{\dag}{\mathbf a}_{j,k}|^2+\sigma^2  \right)-{ r}\alpha\sum_{s=1}^S||{\pmb\phi}_s||_2^2,
\end{aligned}
\end{equation}
and ${\pmb\epsilon}=[\epsilon_1, \ldots, \epsilon_K]^T\in{\mathbb C}^{K\times 1}$ refers to the auxiliary variable vector.
Similarly, we optimize ${\pmb\phi}$ and $\pmb\epsilon$ alternatively \cite{Guo2019Weighted}.
The optimal $\epsilon_k$ for given $\pmb\phi$ can be obtained easily, shown as follows:
\begin{equation}\label{eq:28}
\epsilon_k^{*}=\frac{\sqrt{\tilde{\alpha}_k}(b_{k,k}+{\pmb\phi}^{\dag}{\mathbf a}_{k,k})}
{\sum_{j=1}^K|b_{j,k}+{\pmb\phi}^{\dag}{\mathbf a}_{j,k}|^2+\sigma^2}.
\end{equation}
Then, the remaining problem is optimizing $\pmb\phi$ for given $\pmb\epsilon.$
By introducing new variable ${\pmb\theta}={\pmb\phi}\in{\mathbb C}^{(SN)\times 1}.$
Likewise, ${\pmb\theta}_s\in{\mathbb C}^{N\times 1}$ represents the $s$th block of vector ${\pmb\theta}.$
Thus, for the fixed ${\pmb\epsilon},$ the optimization problem of ${\pmb\phi}$ is given as follows:
\begin{equation}\label{eq:29}
\begin{aligned}
\text{(U2-2)~}\max_{{\pmb\phi}, {\pmb\theta}}~&\sum_{k=1}^K2\sqrt{\tilde{\alpha}_k}\text{Re}\left\{ \epsilon_k^{\ddag}{\pmb\phi}^{\dag}{\mathbf a}_{k,k}+\epsilon_k^{\ddag}b_{k,k}\right\}-\sum_{k=1}^K|\epsilon_k|^2\\
&\left( \sum_{j=1}^K|b_{j,k}+{\pmb\phi}^{\dag}{\mathbf a}_{j,k}|^2+\sigma^2  \right)-{ r}\alpha\sum_{s=1}^S||{\pmb\theta}_s||_2^2\\
\text{s.t.}~& {\pmb\phi}^{\dag}{\mathbf e}_i{\mathbf e}_i^{\dag}{\pmb\phi}\leq 1, \forall i=1, 2, \ldots, (SN)\\
&{\pmb\theta}={\pmb\phi}.
\end{aligned}
\end{equation}

Utilizing the method of augmented Lagrangian minimization, (u2-2) can be handled by solving
\begin{equation}\label{eq:30}
\begin{aligned}
\min_{{\pmb\Lambda}}\max_{{\pmb\phi}, {\pmb\theta}}~&~ L_c({\pmb\phi}, {\pmb\theta}, {\pmb\Lambda})\\
\text{s.t.~}&~{\pmb\phi}^{\dag}{\mathbf e}_i{\mathbf e}_i^{\dag}{\pmb\phi}\leq 1, \forall i=1, 2, \ldots, (SN),
\end{aligned}
\end{equation}
where $c>0$ is the penalty factor; ${\pmb\Lambda}\in{\mathbb C}^{(SN)\times 1}$ is the Lagrangian vector multiplier for ${\pmb\theta}={\pmb\phi}.$
The partial augmented Lagrangian function is defined as
\begin{equation}\label{eq:31}
\begin{aligned}
L_c&({\pmb\phi},{\pmb\theta},{\pmb\Lambda})=\sum_{k=1}^K2\sqrt{\tilde{\alpha}_k}\text{Re}\left\{ \epsilon_k^{\ddag}{\pmb\phi}^{\dag}{\mathbf a}_{k,k}+\epsilon_k^{\ddag}b_{k,k} \right\}\\
&-\sum_{k=1}^K|\epsilon_k|^2\left( \sum_{j=1}^K|b_{j,k}+{\pmb\phi}^{\dag}{\mathbf a}_{j,k}|^2+\sigma^2  \right)\\
&-{ r}\alpha\sum_{s=1}^S||{\pmb\theta}_s||_2^2-
\text{Re}\left\{\text{Tr}\left[{\pmb\Lambda}^{\dag}({\pmb\theta}-{\pmb\phi}) \right]\right\}
-\frac{c}{2}||{\pmb\theta}-{\pmb\phi}||_2^2.
\end{aligned}
\end{equation}
\begin{itemize}
\item{\textit{Updating} ${\pmb\phi}$:

By dual theory and KKT conditions, the optimal solution is given by (\ref{eq:32}).
\begin{figure*}
\begin{equation}\label{eq:32}
\begin{aligned}
{\pmb\phi}^{*}=&\left(2\sum_{k=1}^K|\epsilon_k|^2\left( \sum_{j=1}^K {\mathbf a}_{j,k}{\mathbf a}_{j,k}^{\dag}\right)+2\sum_{i=1}^{SN}\mu_i{\mathbf e}_i{\mathbf e}_i^{\dag}+c{\mathbf I}_{SN} \right)^{-1}\\
&\times\left(2\sum_{k=1}^K\sqrt{\tilde{\alpha}_k}\epsilon_k^{\ddag}{\mathbf a}_{k,k}+{\pmb\Lambda}+c{\pmb\theta}-2\sum_{k=1}^K|\epsilon_k|^2\sum_{j=1}^Kb_{j,k}{\mathbf a}_{j,k}\right),\\
\hline
\end{aligned}
\end{equation}
\end{figure*}
The Lagrangian multiplier $\mu_i$ updated by
\begin{equation}\label{eq:33}
\mu_i^{*}=\max\left\{0, 1-{\pmb\phi}^{\dag}{\mathbf e}_i{\mathbf e}_i^{\dag}{\pmb\phi}\right\}.
\end{equation}
}
\item{\textit{Updating } ${\pmb\theta}$:

The problem of ${\pmb\theta}$ is an unconstrained group  leastabsolute selection and shrinkage operator (group Lasso) problem \cite{Yuan2006Model}, i.e.,
\setcounter{equation}{31}
\begin{equation}\label{eq:34}
\max_{\pmb\theta}~-{\mathbf r}\alpha\sum_{s=1}^S||{\pmb\theta}_s||_2-\text{Re}\left\{\text{Tr}
\left[{\pmb\Lambda}^{\dag}({\pmb\theta}-{\pmb\phi})\right] \right\}-\frac{c}{2}||{\pmb\theta}-{\pmb\phi}||_2^2.
\end{equation}
Let ${\pmb\Lambda}_s\in{\mathbb C}^{N\times 1}$ denote the $s\text{th}$ row block of vector ${\pmb\theta}, s=1, 2, \ldots, S.$
Then, (\ref{eq:34}) can be divided into $S$ independent problems of ${\pmb\theta}_s$ for $s=1, 2, \ldots, S$
\begin{equation}\label{eq:35}
\max_{{\pmb\theta}_s}~-{ r}\alpha||{\pmb\theta}_s||_2-\text{Re}\left\{\text{Tr}\left[{\pmb\Lambda}_s^{\dag}
({\pmb\theta}_s-{\pmb\phi}_s)  \right] \right\}-\frac{c}{2}||{\pmb\theta}_s-{\pmb\phi}_s||_2^2
\end{equation}
Defining ${\mathbf x}_s=c{\pmb\phi}_s-{\pmb\Lambda}_s,$ and ${\mathbf x}_s-c{\pmb\theta}_s\in{ r}\partial ||{\pmb\theta}_s||_2$, and thus,  we can easily obtain
\begin{equation}\label{eq:36}
{\pmb\theta}_s=\left\{ \begin{array}{ccc}
&{\mathbf 0}, & \text{~if~} ||{\mathbf x}_s||_2\leq { r}\\
&\frac{(||{\mathbf x}_s||_2-{\mathbf r}\alpha){\mathbf x}_s}
{c||{\mathbf x}_s||_2}, &\text{otherwise}.
\end{array}\right.
\end{equation}
The update of Lagrangian vector $\pmb\Lambda_s$ is given by
\begin{equation}\label{eq:37}
{\pmb\Lambda}_s={\pmb\Lambda}_s+c({\pmb\theta}_s-{\pmb\phi}_s), \forall s=1, 2, \ldots, S.
\end{equation}
}
\end{itemize}

\subsection{Game Analysis for the IRS Pricing}
Substituting (\ref{eq:36}) into ({\bf L-Problem}) in (\ref{eq:8}), the optimization problem at the IRS side can be formulated as
\begin{equation}\label{eq:38}
\max_{{ r}>0}~ \sum_{s=1}^S\kappa_s\frac{-{ r}^2+||{\mathbf x}_s||_2{ r}}
{c},
\end{equation}
where $\kappa_s$ is indicate function, i.e.,
\begin{equation}\label{eq:39}
\kappa_s=\left\{\begin{array}{ccc}
&0, &\text{~if~} ||{\mathbf x}_s||_2\leq { r}\\
&1, &\text{~otherwise}.
\end{array}\right.
\end{equation}
The optimal solution of (\ref{eq:38}) is
\begin{equation}\label{eq:40}
{ r}^{*}=\frac{\sum_{s=1}^S\kappa_s||{\mathbf x}_s||_2}
{2\sum_{s=1}^S\kappa_s}.
\end{equation}
%The IRS starts with a high price, where no reflection modules would like to be selected to serve the BS. Then, in each round of the circulation, for the IRS, we set small step $\Delta$ and changes its current price ${\mathbf r}$ with $\Delta$ higher or lower than the original price.
%If the utility is the highest when the price increases with $\Delta$, in the next round, the price changes to be ${\mathbf r}+\Delta.$
%If the utility is the highest when the price decreases with $\Delta,$ in the next round, the price changes to be ${\mathbf r}-\Delta.$
%Otherwise, the price remains unchanged. The circulation continuous until the IRS can not deviate from the current price for higher utility.

The entire framework including the identifying the price and the trigger module subsets as well as the transmit beamforming and the phase shift is summarized in Algorithm 1.
\begin{algorithm}[!htp]%\label{algorithm:1}         %算法的开始
\caption{ Algorithm Summary}             % 算法的标题
\label{alg:Framwork}
\hspace*{0.02in} {\textit {Step 0: }} The IRS initialize the price ${\mathbf r}(1),$ and set the outer iteration number $\tau=1$   \\
\hspace*{0.02in} {\bf Part I:} the alternating optimization for solving (U0-1) \\
(1.1) Initialize ${\mathbf W}(1)$ and ${\pmb\Phi}(1)$ to feasible values, and set the iteration number $t=1.$\\
\hspace*{0.02in} {\bf Repeat}\\
(1.2) Update the nominal SINR $\alpha_k(t), \forall k\in{\cal K},$ by (\ref{eq:13});\\
(1.3) Update $\beta_k(t), \forall k\in{\cal K}$ by (\ref{eq:21});\\
(1.4) Update transmit beamforming ${\mathbf W}(t)$ by (\ref{eq:22}); update $\lambda_0(t)$ by (\ref{eq:23})\\
(1.5) Update $\epsilon_k(t), \forall k\in{\cal K}$ by (\ref{eq:28});
(1.6) Update ${\pmb\phi}(t)$ by (\ref{eq:32}); update $\mu_i(t), \forall i=1, 2, \ldots, (SN),$ by (\ref{eq:33});
(1.7) Update ${\pmb\theta}_s(t) $ by (\ref{eq:36}) in parallel for $s=1, 2,\ldots S;$\\
(1.8) Update $\pmb\Lambda_s(t)$ by (\ref{eq:37}) in parallel for $s=1, 2,\ldots, S;$\\
(1.9) Update $t=t+1;$
(1.10) {\bf Until } The value of function (\ref{eq:12}) converges.\\
\hspace*{0.02in} {\bf Part II:} Update price ${\mathbf r}$ by solving problem (\ref{eq:38}) in the outer loop \\
(2.1) Solve problem (\ref{eq:38}) for given $\{{\pmb\theta}_s(t)\}_{s=1}^S, \{{\pmb\Lambda}_s(t)\}_{s=1}^S, $  update ${ r}(\tau)$ by (\ref{eq:40})\\
(2.2) {\bf Until } the utility of the IRS is convergence.
\end{algorithm}

\section{Simulation Results}
\begin{figure}[!t]
\centering
\begin{minipage}[t]{0.45\textwidth}
\centering
\includegraphics[width=6.5cm]{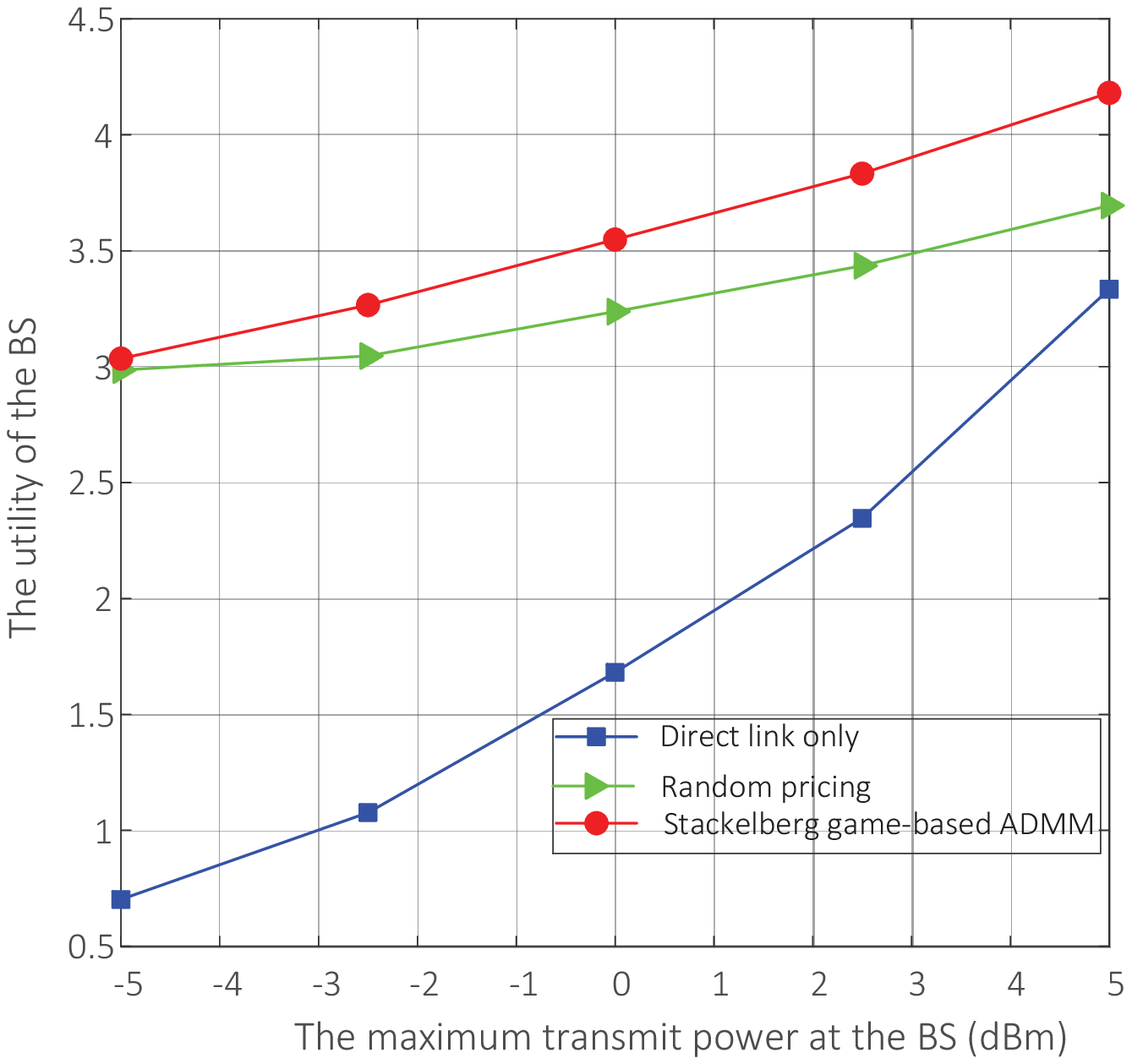}
\caption{Impact of the maximum transmit power at the BS on the utility of the follower, i.e., the BS.}
\label{fig:1}
\end{minipage}
\begin{minipage}[t]{0.45\textwidth}
\centering
\includegraphics[width=6.5cm]{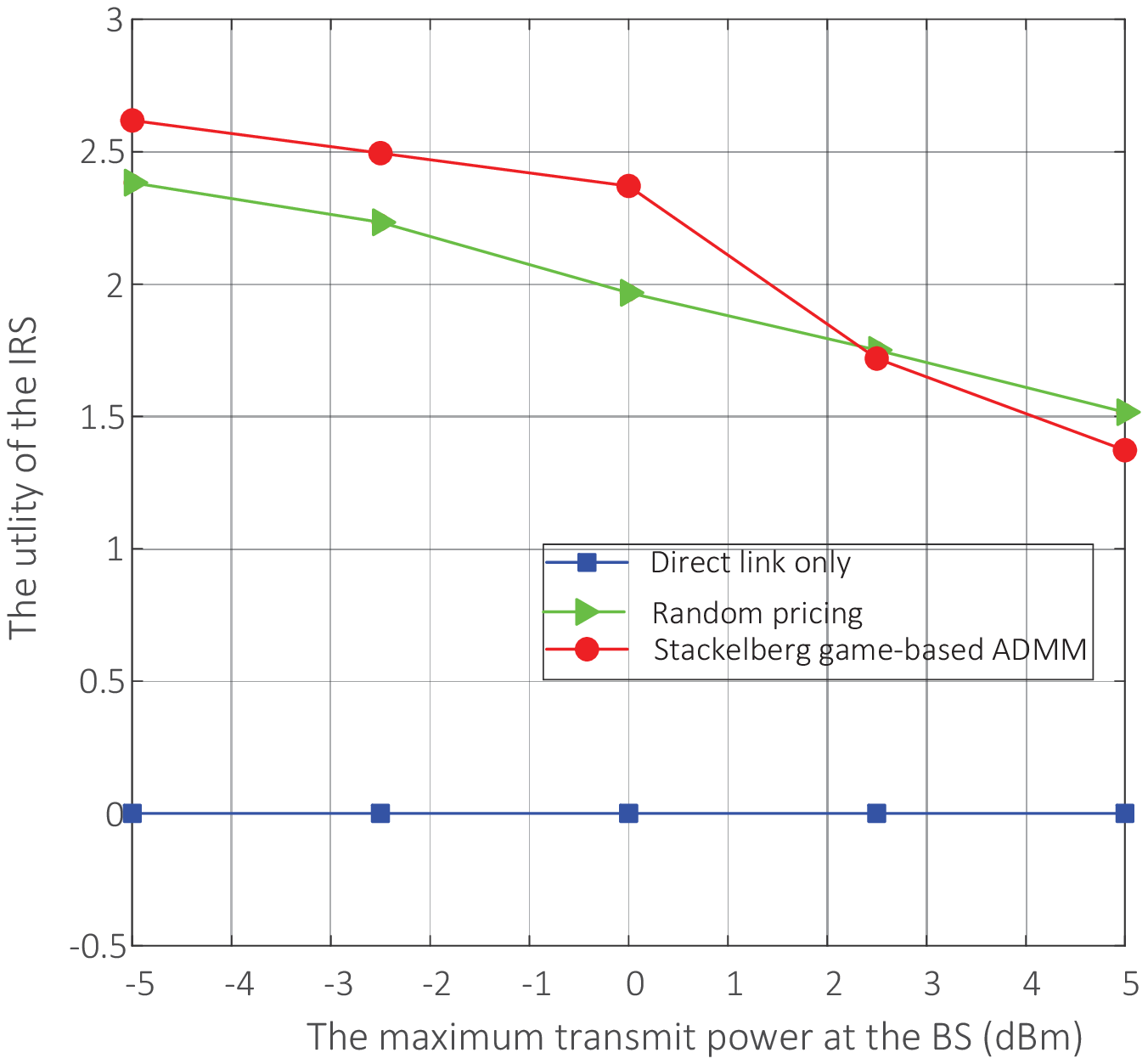}
\caption{Impact of the maximum transmit power at the BS on the utility of the leader, i.e., the IRS.}
\label{fig:2}
\end{minipage}
\end{figure}
\begin{figure}[!t]
\centering
\begin{minipage}[t]{0.45\textwidth}
\centering
\includegraphics[width=6.5cm]{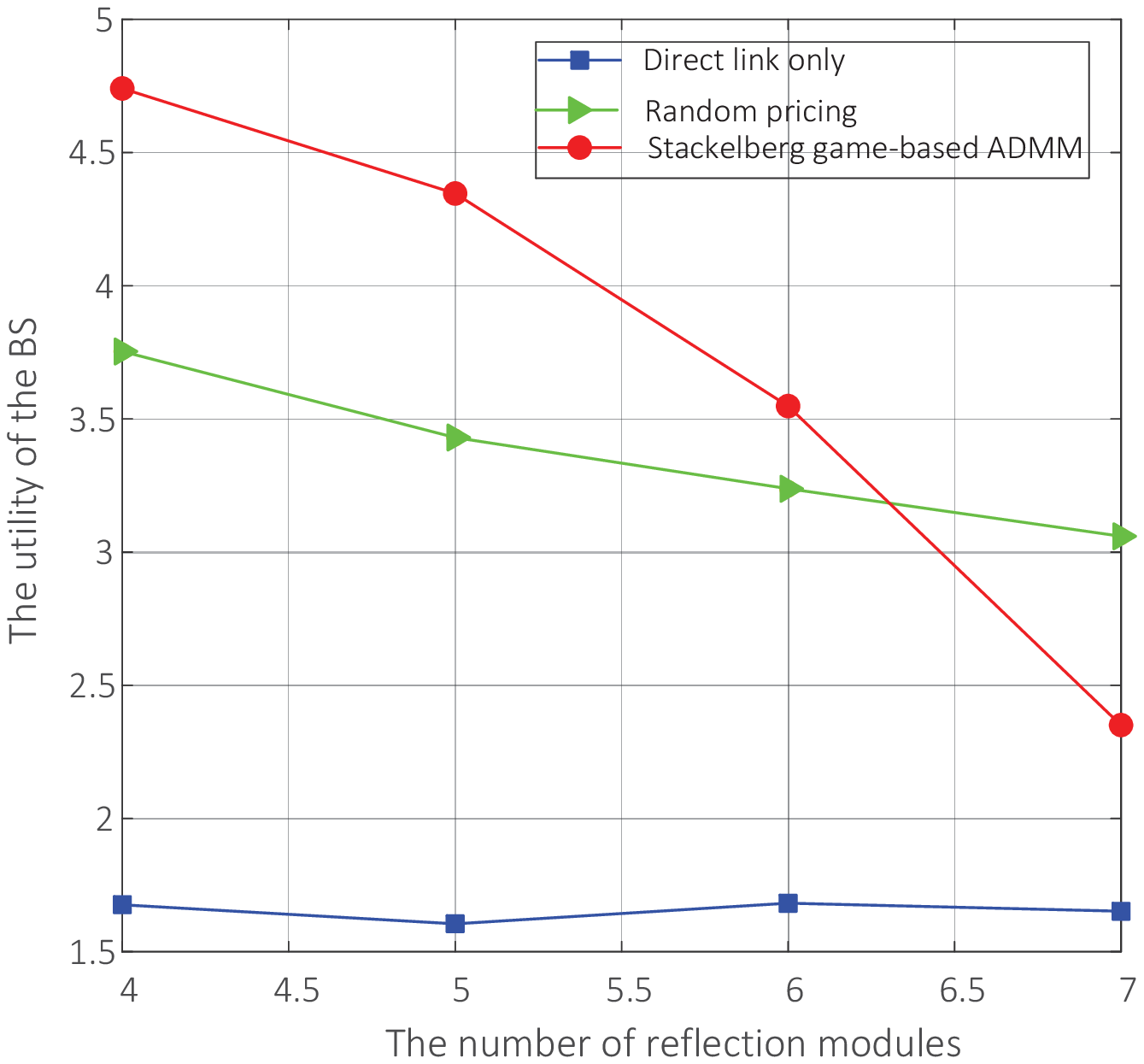}
\caption{Impact of the number of reflection modules of the IRS on the utility of the follower, i.e., the BS, when each module consisting $8$ reflection elements. }
\label{fig:3}
\end{minipage}
\begin{minipage}[t]{0.45\textwidth}
\centering
\includegraphics[width=6.5cm]{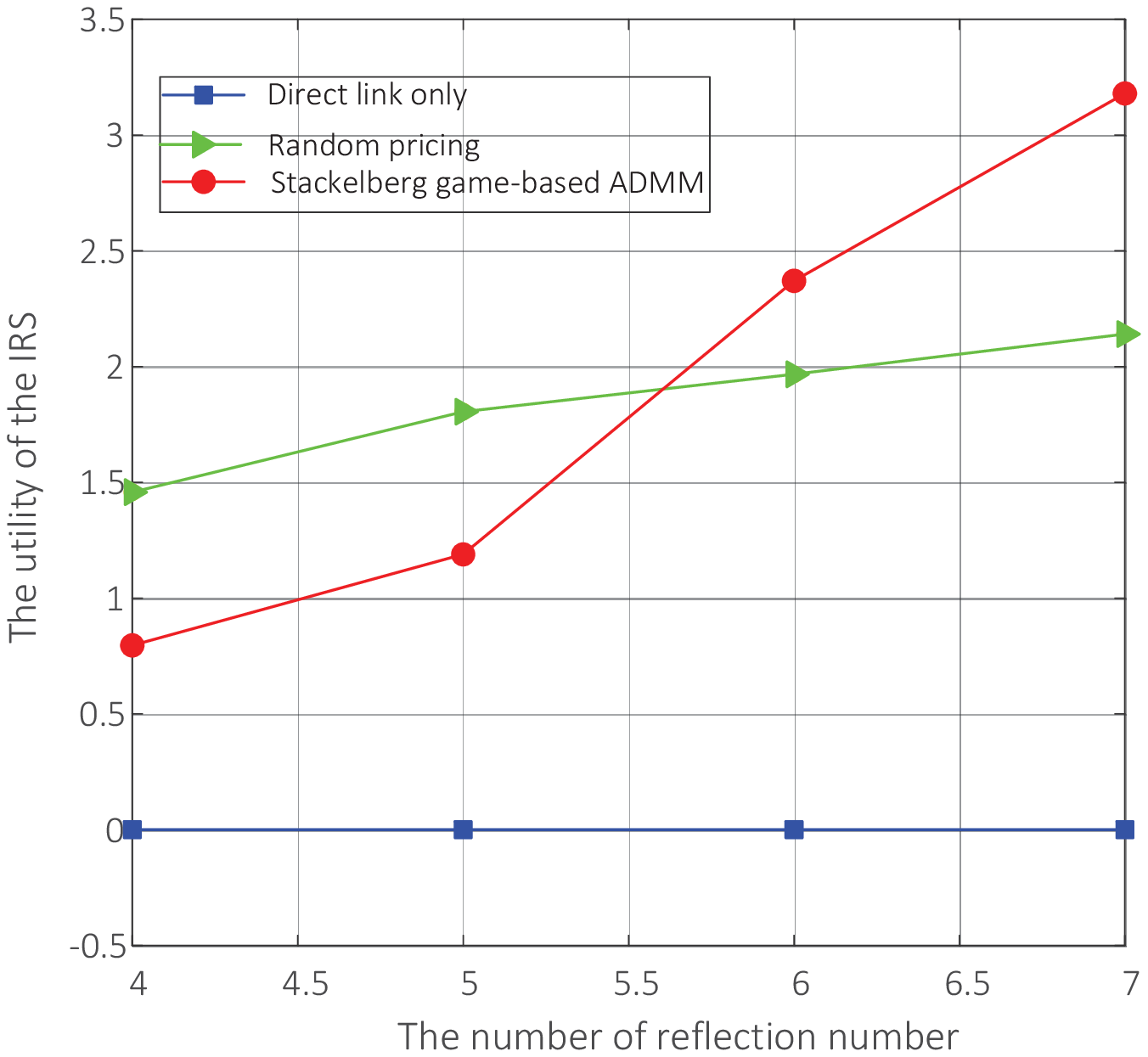}
\caption{Impact of the number of reflection modules of the IRS on the utility of the leader, i.e., the IRS, when each module consisting $8$ reflection elements.}
\label{fig:4}
\end{minipage}
\end{figure}
In this section, extensive numerical results are presented to evaluate the performances of the proposed resource allocation strategies based on the approach of  trigger module pricing. For simplicity, we set the balance parameter $\alpha$ to be  $0.1.$
The comparison results between the proposed Stackelberg game-based ADMM scheme, the random pricing, and direct link only scheme is presented to demonstrate the higher utility of the BS and IRS in the proposed scheme. Then the comparative summary between the Stackelberg game-based ADMM and some existing schemes are presented.
To keep the complexity of the simulations tractable, we focus on the scenario, where the $K=4$ users are randomly employed within a circle cell centered at $(200,0)~\text{m}$, and the cell radius is $10~\text{m}$, the BS and IRS are employed at $(0,0)~\text{m} $ and $(200, 50)~\text{m},$ respectively,  where the number of reflection elements of each module is set as $N=8.$
We assume that the BS is equipped with $4$ antennas.
We assume quasi-static block fading channels, i.e., the channels from the BS to the IRS and the IRS to the users remain constant during each time block, but may vary from one to another \cite{Yang2019Low}.
To include the effects of fading and shadowing, we use the path-loss model introduced in \cite{Gao2019Dynamic}.

The performance of the Stackelberg game-based ADMM scheme is evaluated against two existing benchmark schemes, i.e., random pricing scheme and direct link only scheme in Fig. \ref{fig:1} and Fig. \ref{fig:2}.
In the random pricing scheme, the IRS randomly determines its strategies, without considering the existence of the BS.
The direct link only scheme  means no IRS to aid, i.e., no module is triggered at IRS.
Fig. \ref{fig:1} and Fig. \ref{fig:2} respectively show the effect of the maximum transmit power $P^{\max}$ on the utility of the BS and the IRS, when the number of reflection modules is $6.$
For the BS and IRS, the Stackelberg game-based ADMM scheme achieves the highest utility value compared with random pricing and direct link schemes, which indicates that the proposed pricing-based Stackelberg game scheme performs best in resource allocation for IRS-aided communications.
From the results, we observe that the utility values of the BS increases as $p^{\max}$ grows from $-5\text{~dBm}$ to $5\text{~dBm}.$
Meanwhile,  the utility value of the IRS achieved by the Stackelberg game-based ADMM scheme first  decreases slowly until the maximum transmit power increases to $0\text{~dBm}$ and then decreases rapidly by increasing the value of $p^{\max}$.
This is because that the cost of power consumption is not considered in the utility of the BS, and thereby,  the BS will tend to select a small number of reflection modules when the transmit power is sufficient.

Fig. \ref{fig:3} and Fig. \ref{fig:4} illustrate how the number of reflection modules $S$ affects the utility values of the BS and IRS, respectively, when the maximum transmit power $p^{\max}=0\text{dBm}.$
The IRS's utility values in the Stackelberg game scheme and the random pricing scheme increase as the number of reflection modules grows from $S=4$ to $S=7.$
This is due to the fact that the IRS operator incentivizes the BS to trigger more reflection modules by appropriately adjusting pricing strategies.
Consequently, the utility values of the BS decrease as the number of reflection modules, $S$, increases.
Most importantly, Figs. \ref{fig:1}--\ref{fig:4} show that the proposed Stackelberg game-based ADMM scheme outperforms the other two schemes in pricing-based resource allocation.

\section{Conclusion}
The adoption of an IRS for downlink multi-user communication from a multi-antenna BS was investigated in this paper.
%Nevertheless, in a real-world network, the IRS operator may be selfish or has its own objectives due to competing/limited resources as well as deployment/maintenance cost.
Specifically,  we developed a Stackelbeg game approach to analyze the interaction between the BS and the IRS operator considering that the IRS operator may be selfish or has its own objectives.
Different from the existing studies on IRS that merely focused on tuning the reflection coefficient of all the reflection elements, we considered the  reflection resource (elements) management, which can be realized via trigger module selection under our proposed IRS architecture that all the reflection elements are partially controlled by independent switches of controller.
A Stackelberg game-based ADMM was proposed to solve either the transmit beamforming at the BS or the passive beamorming of the triggered reflection modules.
Numerical examples were presented to verify the proposed studies.
It was shown that the proposed scheme is effective in the utilities of both the BS and IRS.

\bibliographystyle{ieeetr}
\bibliography{newslsf}

\begin{thebibliography}{10}

\bibitem{Wu2019Beamforming}
Q.~Wu and R.~Zhang, ``Beamforming optimization for wireless network aided by
  intelligent reflecting surface with discrete phase shifts,'' pp.~1--30, June
  2019.

\bibitem{Huang2019Reconfigurable}
C.~Huang, A.~Zappone, G.~C. Alexandropoulos, M.~Debbah, and C.~Yuen,
  ``Reconfigurable intelligent surfaces for energy efficiency in wireless
  communication,'' {\em IEEE Transactions on Wireless Communications}, vol.~18,
  pp.~4157--4170, August.

\bibitem{Zenzo2019Wireless}
E.~Basar, M.~D. Renzo, J.~D. Rosny, M.~Debbah, M.~S. Alouini, and R.~Zhang,
  ``Wireless communications through reconfigurable intelligent surfaces,'' {\em
  IEEE Access}, vol.~7, pp.~116753--116773, 2019.

\bibitem{Zenzo2019Smart}
M.~D. Renzo, M.~Debbah, D.~T. Phan-Huy, and A.~Zapppone, ``Smart radio
  environments empowered by reconfigurable {AI} meta-surfaces: an idea whose
  time has come,'' {\em Eurasip Journal on Wireless Communications and
  Networking}, pp.~1--32, May 2019.

\bibitem{Basar2019Wireless}
E.~Basar, M.~D. Renzo, J.~D. Rosny, M.~Debbah, M.~Alouini, and R.~Zhang,
  ``Wireless communications through reconfigurable intelligent surfaces,'' {\em
  IEEE Access}, vol.~7, pp.~116753--116773, August 2019.

\bibitem{Hu2019Reconfigurable}
J.~Hu, H.~Zhang, B.~Di, L.~Li, L.~Song, Y.~Li, Z.~Han, and H.~V. Poor,
  ``Reconfigurable intelligent surfaces based rf sensing: design, optimization,
  and implementation.''
  \url{https://ui.adsabs.harvard.edu/abs/2019arXiv191209198H/abstract}, 2019.
\newblock Online.

\bibitem{Wu2019Towards}
Q.~Wu and R.~Zhang, ``Towards smart and reconfigurable environment: intellignet
  reflecting surfaces aided wireless network,'' 2019.

\bibitem{Han2019Large}
Y.~Han, W.~Tang, S.~Jin, C.~Wen, and X.~Ma, ``Large intelligent surfaceassisted
  wireless communication exploiting statistical {CSI},'' {\em IEEE Trans. Veh.
  Technol.}, vol.~68, no.~8, pp.~8238--8242, 2019.

\bibitem{Gao2020Reconfigurable}
Y.~Gao, C.~Yong, Z.~H. Xiong, D.~Niyato, and Y.~Xiao, ``Reflection resource
  management for intelligent reflecting surface aided wireless networks.''
  \url{https://arxiv.org/abs/2002.00331}, 2020.
\newblock Online.

\bibitem{Wu2019Intelligentjournal}
Q.~Wu and R.~Zhang, ``Intelligent reflecting surface enhanced wireless network
  via joint active and passive beamforming design,'' {\em IEEE Transactions on
  Wireless Communications}, August 2019.

\bibitem{Guo2019Weighted}
H.~Guo, Y.~C. Liang, J.~Chen, and E.~G. Larsson, ``Weighted sum-rate
  optimization for intelligent reflecting surface enhanced wireless networks,''
  2019.

\bibitem{Pan2019Intelligent}
C.~Pan, H.~Ren, K.~Wang, M.~Elkashlan, A.~Nallanathan, J.~Wnag, and L.~Hanzo,
  ``Intelligent reflecting surface aided mimo broadcasting for simultaneous
  wireless information and power transfer,'' 2019.

\bibitem{Bjornson2019Intelligent}
E.~Bjornson, O.~Ozdogan, and E.~G. Larsson, ``Intelligent reflecting surface
  vs. decode-and-forward: How large surfaces are needed to beat relaying?,''
  pp.~1--5, August 2019.

\bibitem{Zenzo2019Reconfigurable}
K.~Ntontin, M.~D. Renzo, J.~Song, F.~Lazarakis, J.~D. Rosny, D.-T. Phan-Huy,
  O.~Simeone, R.~Zhang, M.~Debbah, G.~Lerosey, M.~Fink, S.~Tretyakov, and
  S.~Shamai, ``Reconfigurable intelligent surfaces vs. relaying: Differences,
  similarities, and performance comparison.''
  \url{https://arxiv.org/abs/1908.08747}, 2019.
\newblock Online.

\bibitem{Liu2019Intelligent}
F.~Liu, O.~Tsilipakos, A.~Pitilakis, A.~C. Tasolamprou, M.~S. Mirmoosa, N.~V.
  Kantartzis, and et. al., ``Intelligent metasurfaces with continuously tunable
  local surface impedance for multiple reconfigurable functions,'' {\em
  Physical Review Applied}, vol.~11, pp.~044024--1--044024--1, April 2019.

\bibitem{Tan2018Enabling}
X.~Tan, Z.~Sun, D.~Koutsonikolas, and J.~M. Jornet, ``Enabling indoor mobile
  millimeter-wave networks based on smart reflect-arrays,'' in {\em 2018 IEEE
  Conference on Computer Communications}, (Honolulu, HI, USA), pp.~1--9, April
  2018.

\bibitem{Li2017Electromagnetic}
L.~Li, C.~T. Jun, W.~Ji, S.~Liu, J.~Ding, X.~Wan, L.~Y. Bo, M.~Jiang, C.~Qiu,
  and S.~Zhang, ``Electromagnetic reprogrammable coding-metasurface
  holograms,'' {\em Nature Communications}, vol.~8, pp.~1--7, August 2017.

\bibitem{Mehanna2013Joint}
O.~Mehanna, N.~D. Sidiropoulos, and G.~B. Giannakis, ``Joint multicast
  beamforming and antenna selection,'' {\em IEEE Transactions on Signal
  Processing}, vol.~61, pp.~2660--2674, May 2013.

\bibitem{Yuan2006Model}
M.~Yuan and Y.~Lin, ``Model selection and estimation in regression with grouped
  variables,'' {\em Journal Royal Statistical Society}, vol.~68, pp.~49--67,
  December 2006.

\bibitem{Yang2019Low}
Z.~Yang and M.~Dong, ``Low-complexity coordinated relay beamforming design for
  multi-cluster relay interference networks,'' {\em IEEE Transactions on
  Wireless Communications}, vol.~18, no.~4, pp.~2215--2228, 2019.

\bibitem{Gao2019Dynamic}
Y.~Gao, Y.~Xiao, M.~Wu, M.~Xiao, and J.~Shao, ``Dynamic social-aware peer
  selection for cooperative relay management with d2d communications,'' {\em
  IEEE Transactions on Communications}, vol.~67, pp.~3124--3139, May 2019.

\end{thebibliography}

\end{document}